\documentclass[useAMS,usenatbib]{mn2e}

\title[]{The role of thermal conduction in magnetized viscous-resistive ADAFs}
\author[J. Ghanbari , S. Abbassi  and M. Ghasemnezhad]{J.
Ghanbari$^{1}$\thanks{E-mail: ghanbari@ferdowsi.um.ac.ir}
, S. Abbassi $^{2,3}$\thanks{E-mail:sabbassi@dubs.ac.ir} and M. Ghasemnezhad$^1$ \\
$^{1}$Department of Physics, School of Sciences, Ferdowsi University of Mashhad, Mashhad, 91775-1436, Iran\\
$^{2}$School of Physics, Damghan University of Basic Sciences, P.O.Box 36715-364, Damghan, Iran\\
$^{3}$School of Astronomy, Institute for Research in Fundamental Sciences (IPM), P.O.Box 19395-5531, Tehran, Iran}
\begin{document}
\date{}

\pagerange{\pageref{firstpage}--\pageref{lastpage}} \pubyear{2004}

\maketitle \label{firstpage}

\begin{abstract}
Observations of the hot gas, which is surrounding Sgr A* and a few other
nearby galactic nuclei, imply that mean free paths of electron and
proton are comparable to gas capture radius. So, hot accretion
flows likely proceed under week collision conditions. As a result
thermal conduction by ions has a considerable contribution in
transfer of the realized heat in accretion mechanisms. We
study a 2D advective accretion disk bathed in a poloidal magnetic
field of a central accretor in the presence of thermal
conduction. We find self-similar solutions for an  axisymmetric,
rotating, steady, viscose-resistive, magnetized accretion flow. The
dominant mechanism of energy dissipation is assumed to be
turbulence viscosity and magnetic diffusivity due to magnetic
field of the central accretor. We show that the global structure
of ADAF s are sensitive to viscosity, advection and thermal
conduction parameters. We discuss how radial flow, angular
velocity and density of accretion flows may vary with the advection,
thermal conduction and viscous parameters.

\end{abstract}


\section{INTRODUCTION}

The foundations of our present understanding about advection
dominated accretion flows were laid out in a series of
papers by Narayan $\&$ Yi (1994,1995 a,b), although some ideas
were anticipated much earlier by Ichimaro (1977). The specific
abbreviation ADAF , which stands for advection dominated accretion
flow , was introduced by Lasota et al. (1996). An ADAF is defined as one
in which a large fraction of the viscously generated
heat is advected with the accreting gas, and only a small fraction
of the energy is radiated. ADAFs have an opposite regime in comparison with the
standard model. In the standard model, flow is described in away that the heat
generated by the viscosityradiates out of the system immediately after its generation (Shakura $\&$ Sunyaev 1973).

These advection-dominated accretion flows occur in two regimes
depending on their mass accretion and optical depth. Actually the optical depth of accretion flows is
highly dependent on accretion rates. In a high mass accretion
rate, the optical depth becomes very high and the radiation
generated by the accretion flow can be trapped within the disk.
This type of accretion disks is known as optically
thick or Slim Disks which has been introduced by
Abramowicz et al. (1998). In the limit of low mass accretion rate,
the disk becomes optically thin. In this case, the cooling time of
accretion flows is longer than accreting time scale. So the
energy generated by accretion flows mostly remains in the disks
and the disks can not radiate their energies efficiently. This kind of
accretion flows are named radiation inefficient accretion
flows (RIAFs). This type of accretion flows is investigated by many
authors (Narayan \& Yi 1994, Abramowicz et al. 1995, Chen 1995).

These types of solutions have been used to interpret the
spectra of  X-ray binary black holes in their quiescent or
low/hard state as an alternative to the Shapiro, Lightman and
Eardly (1976, SLE) solutions. Since ADAFs have large radial
velocities and also the infalling matter caries the thermal energy to the
black hole, the energy transported by advection can stabilize the thermal
instability by removing their steep temperature gradient; thus the ADAF
models have been widely used to explain the observations of low luminous
observed in Sgr A* (Narayan et al. 1996, Hameuray et al. 1997).
However, numerical simulations of radiation inefficient accretion
flows revealed that the low viscous flows are convectively unstable
and therefore convection strongly influences the global structure of accretion flows
(Igumenshchev, Abramowicz \& Narayan 2000). Thus, another type of
accretion flows was proposed, in which the convection plays a
dominat mechanism in transporting the energy, angular
momentum and local released viscose energy within the disk.

A remarkable problem arises when the accretion disks
threaded by a magnetic field. In the ADAF models, the temperature
of accreting disks is so high that the accreting materials are
ionized. So, the magnetic field plays an important role in
the dynamics of the accretion flows. Some authors
tried to solve the MHD equations of magnetized ADAFs analytically.
For example, Kaburaki (2000) has presented a set of analytical
solutions for a fully advective accretion flow in a
global magnetic field. Shadmehri (2004) has extended this
analysis for a non-constant resistivity. Ghanbari et al. (2007)
have presented a set of self-similar solutions for 2D
viscous-resistive advection dominated accretion flows (ADAFs) in
the presence of dipolar magnetic field of the central accretor.
They have shown that the presence of magnetic field and its
associated resistivity can considerably change the picture of the accretion
flows.

Recent observations of hot accretion flows around
active galactic nuclei indicate that they should be based on the
collisionless regimes.

Chandra observations provide tight constraints on both  density and
temperature of gas at or near the Bondi capture radius in Sgr A* and several
other nearby galactic nuclei. Tanaka \& Menou (2006) with some calculations have shown that
the accretion disks in such systems will proceed under weakly-collisional conditions.
So, the thermal conduction has an important role in the energy transport along the disks.
The aim of this work is to consider the effect of thermal
conduction which has been largely neglected before, as an energy transport mechanism,
on the 2D structure of ADAFs. It could affect the global properties of hot accretion flows
substantially. A few authors considered the role of
turbulent heat transport in ADAF disks (Honma 1996 , Manmoto et
al.2000). Since thermal conduction acts to oppose the
formation of temperature gradient that causes it, one might
expect that the temperature and density profiles for accretion
flows are modified in which thermal conduction plays a significant role to
appear different, compared to those flows which thermal
conduction is less effective (Shadmehri 2008).

The weakly-collisions nature of hot accretion flows has
been addressed previously (Mahadevan \& Quataret 1997). Johanson
\& Quataret (2007) studied the effect of electron thermal
conduction on the properties of hot accretion flows under the
assumption of spherical symmetry. In another interesting
analysis, Tanaka $\&$ Menou (2006), studied the effect of
saturated thermal conduction on optically thin ADAFs using an
extension of self-similar solution of Narayan \& Yi (1994). In their
solutions, the thermal conduction is provided an extra degree of
freedom which affects the global dynamical behaviors of the
accretion flow. Abbassi et al. (2008) have presented a set of
self-similar solutions for ADAFs with a toroidal magnetic field in
which the saturated thermal conduction has a great role in the
energy transport in the radial direction. The tangled magnetic
field in accretion flows would likely reduce the effective mean
free paths of particles. The magnitude of this reduction which depends on the magnetic field geometry, is still unknown.
We have accounted this possibility by allowing the value of saturated constant,
$\phi_s$, to vary in our solutions. Magnetic field also has
an important role for transferring angular momentum along the disks. So the dynamical
structure of the disk will be affected by magnetic field strength and configuration. So investigating
the magnetized accretion flow with thermal conduction is an important
issue.

\section{The Basic Equations}
We describe the 2-D hot accretion flow with similar maner
of Narayan \& Yi (1995). We adopt spherical polar coordinates ($r,
\theta, \phi$) for axi-symmetric and steady state flows
($\frac{\partial}{\partial \phi}=\frac{\partial}{\partial
t}=0$).The fundamental MHD governing equations can be written as:

The equation of continuity gives

\begin{equation}
\frac{D\rho}{Dt}+\rho\nabla\cdot\textbf{u}=0
\end{equation}

The equation of motion
\begin{equation}
\rho\frac{D\textbf{u}}{Dt}=-\nabla\textbf{P}-
\rho\nabla\phi-\mu\nabla^2\textbf{u}+
(\mu_{b}+\frac{1}{3}\mu)\nabla(\nabla\cdot\textbf{u})+
\frac{1}{4\pi}\textbf{J}\times\textbf{B}
\end{equation}

The equation of energy
\begin{equation}
\rho[\frac{D\varepsilon}{Dt}+P\frac{D}{Dt}
(\frac{1}{\rho})]=Q_{vis}+ Q_{B}-Q_{rad}+Q_{cond}
\end{equation}

Gauss's law
\begin{equation}
\nabla\cdot\textbf{B}=0
\end {equation}

and the induction equation
\begin{equation}
\frac{DB}{Dt}=\nabla\times(\textbf{u}\times\textbf{B})+\eta\nabla^2\textbf{B}
\end{equation}
where $\rho$ is the density of the gas, p the pressure , $\epsilon$
the internal energy, u is the flow velocity, B is the
 the magnetic field, $\textbf{J}=\nabla\times\textbf{B}$ the current density, $\eta$
the magnetic diffusivity in which for simplicity it is assumed to
be a constant parameter (see, e.g., Kaburaki 2000), $\mu$ and
$\mu_b$ are the shear and bulk viscosities .

The viscous heating rate is defined as the expression:
\begin{equation}
Q_{vis}=2\mu
E_{ij}E^{ij}+(\mu_b-\frac{2}{3}\mu)({\nabla}\cdot\textbf{v})^2\
\end{equation}
where $E_{ij}=\frac{1}{2}(v_{i,j}+v_{j,i})$ is a symmetric tensor
and is known as the rate of the strain tensor.

We have adopted saturated conduction (Cowie $\&$Mckee 1977)
as:
\begin{equation}
Q_{cond}=-\nabla\cdot F_{s}
\end{equation}
where as we have already mentioned $F_{s}=5\phi_{s}\rho c_s^3$ is the
saturated conduction flux on the direction of the temperature
gradient. Tanka \& Menou (2006) have shown that for very small $\phi_s$ their solutions
coincide the standard ADAF solutions.  \textbf{They have shown that}
 by adding the saturated conduction parameter
, $\phi_s$, the effect of thermal conduction can be better seen when we will
approach to $\sim0.001-0.01$. So, we have investigated the effect of thermal conduction in this range.

So, magnetic reconnection may lead to energy release. Also,we can
consider the viscous and resistive dissipations due to a turbulence
cascade. In this study,the resistive dissipation is defined:
\begin{equation}
Q_{B}=\frac{\eta}{4\pi}J^2
\end{equation}

In the right hand side of the energy equation we have:
\begin{displaymath}
Q_{+}-Q_{-}+Q_{cond}=Q_{adv}=fQ_{+}+Q_{cond}
\end{displaymath}
where $Q_{+}=Q_{vis}+Q_{B}, Q_{-}=Q_{rad}$ and $Q_{adv}$
represents the advective transport of energy and is defined as
the difference between the magneto-viscous heating rate, $Q_{+}$,
and radiative cooling rate, $Q_{rad}$ plus the energy transport
by conduction, $Q_{cond}$.  We employ the parameter
$f=1-\frac{Q_{-}}{Q_{+}}$ to measure the hight degree to which
accretion flow is advection-dominated. When $f\sim 1$ the
radiation can be neglected and the accretion flow is advection
dominated while in the case of small $f$  the disk is in the
radiation dominated case. So we can rearrange the right hand side
of the energy equation to $fQ_{+}+Q_{cond}$, where $f\leq 1$. In
general, it varies with $r$ and depends on the details of heating
and cooling processes. For simplicity, it is assumed to be a constant.

For simplicity,the self-gravity of the disc and the effect of
general relativity have been neglected. Also, we neglect radiation
pressure in the equations because in optically thin ADAFs,
$P^{gas}\gg P^{rad}$. We adopt the dipolar configuration
for the magnetic field. Also we have neglected the
$\theta$-component of the flow velocity $u_{\theta}=0$,and the
bulk viscosity of the flow, $ \mu_{b}=0$. Now , we formulate the
basic equations (1)-(5) in spherical polar coordinates as
follows:
\begin{equation}
\frac{\partial\rho}{\partial
t}+\frac{1}{r^2}\frac{\partial}{\partial r}(r^2\rho
u_r)+\frac{1}{r}\frac{\partial}{\partial\theta}(\rho
u_\theta)=0\label{con2},
\end{equation}
The three components of the momentum equations give (e.g.,
Mihalas $\&$ Mihalas 1984):
r component

\begin{displaymath}
\rho[u_{r}\frac{\partial u_{r}}{\partial r}-
\frac{u_{\varphi}^2}{r}] =-\frac{GM\rho}{r^2}-\frac{\partial p
}{\partial r}+\mu[\frac{4}{3}\frac{\partial^2 u_{r}}{\partial
r^2}+\frac{8}{3}\frac{1}{r}\frac{\partial u_{r}}{\partial r}
\end{displaymath}
\begin{displaymath}
-\frac{8}{3}\frac{u_{r}}{r^2}+ \frac{1}{r^2} \cot\theta
\frac{\partial u_{r}}{\partial\theta}
+\frac{1}{r^2}\frac{\partial}{\partial\theta}(\frac{\partial
u_{r}}{\partial\theta})]
\end{displaymath}
\begin{equation}
+\frac{1}{4\pi}[-\frac{B_{\theta}}{r}(\frac{\partial}{\partial
r}(r B_{\theta})-\frac{\partial B_{r}}{\partial\theta})-
\frac{B_{\varphi}}{r}\frac{\partial}{\partial r}(r B_{\varphi})]
\end{equation}

$\theta$ component

\begin{displaymath}
\rho[-\frac{cot\theta}{r}u_{\varphi}^2]=-\frac{1}{r}\frac{\partial
P}{\partial\theta}+\mu[\frac{8}{3}\frac{1}{r^2}\frac{\partial
u_{r}}{\partial\theta}+\frac{1}{3}\frac{1}{r}\frac{\partial^2
u_{r}}{\partial r\partial\theta}]
\end{displaymath}
\begin{equation}
+\frac{1}{4\pi}[\frac{B_{r}}{r}(\frac{\partial}{\partial
r}(rB_{\theta})-\frac{\partial
B_{r}}{\partial\theta})-\frac{B_{\varphi}}{r
\sin\theta}\frac{\partial}{\partial\varphi}(B_{\varphi}\sin\theta)]
\end{equation}

$\varphi$ component

\begin{displaymath}
\rho[u_{r}\frac{\partial u_{\varphi}}{\partial
r}+\frac{u_{r}u_{\varphi}}{r}]=\mu[\frac{\partial^2
u_{\varphi}}{\partial r^2}+\frac{2}{r}\frac{\partial
u_{\varphi}}{\partial r}+\frac{1}{r^2}cot\theta\frac{\partial
u_{\varphi}}{\partial\theta}
\end{displaymath}
\begin{displaymath}
+\frac{1}{r^2}\frac{\partial^2
u_{\varphi}}{\partial\theta^2}-\frac{u_{\varphi}}{r^2
\sin^2\theta}]+\frac{1}{4\pi}[B_{\theta}(\frac{1}{r
\sin\theta}\frac{\partial}{\partial\theta}(\sin\theta
B_{\varphi})
\end{displaymath}
\begin{equation}
+\frac{B_{r}}{r}\frac{\partial}{\partial r}(r
B_{\varphi})]
\end{equation}
the equation of energy
\begin{displaymath}
\rho[u_{r}\frac{\partial\varepsilon}{\partial r}-\frac{P
u_{r}}{\rho^2}\frac{\partial\rho}{\partial r}]=-\frac{2}{3}\mu
f[\frac{1}{r^2}\frac{\partial}{\partial r}(r^2 u_{r})]^2+2\mu
f[(\frac{\partial u_{r}}{\partial r})^2
\end{displaymath}
\begin{displaymath}
+2(\frac{u_{r}}{r})^2 +\frac{1}{2}( \frac{1}{r}\frac{\partial
u_{r}}{\partial\theta})^2 +\frac{1}{2}[r\frac{\partial}{\partial
r}(\frac{u_{r}}{r})]^2+\frac{1}{2}[\frac{\sin\theta}{r}\frac{\partial}
{\partial\theta}(\frac{u_{\varphi}}{\sin\theta})]^2
\end{displaymath}
\begin{displaymath}
]+\frac{\eta}{4\pi r^2}[\frac{\partial}{\partial r} (r
B_{\theta})-\frac{\partial
B_{r}}{\partial\theta}]^2-\frac{1}{r^2}\frac{\partial}{\partial
r} (5 \Phi_{s} r^2 P^{3/2} \rho^{-1/2})
\end{displaymath}
\begin{equation}
-\frac{1}{r sin\theta}\frac{\partial}{\partial\theta}
(5\Phi_{s}sin\theta P^{3/2}\rho^{-1/2})
\end{equation}

The three components of the induction equation
\begin{equation}
u_{r}\frac{\partial B_{r}}{\partial
r}=\frac{\partial}{\partial\theta}[r
\sin\theta(u_{r}B_{\theta}-\frac{\eta}{r}[\frac{\partial}{\partial
r}(r B_{\theta})-\frac{\partial B_{r}}{\partial\theta}])]
\end{equation}

\begin{equation}
-\frac{1}{r \sin\theta}(\frac{\partial}{\partial r}(r
\sin\theta[u_{r}B_{\theta}-\frac{\eta}{r}(\frac{\partial}{\partial
r}(r B_{\theta})-\frac{\partial B_{r}}{\partial\theta})]))=0
\end{equation}

\begin{equation}
\frac{1}{r}[\frac{\partial}{\partial r}(r u_{\varphi}
B_{r})+\frac{\partial}{\partial\theta}(u_{\varphi} B_{\theta})]=0
\end{equation}

Now we have a set of MHD equations which describe the
dynamical behavior of magnetized ADAFs. The solution of these
equations are strongly depends on viscosity, resistivity, degree
of advection and the role of thermal conduction on the disks.

These nine partial differential equations governing the non-self
gravitating, magnetized advection dominated viscose
flows. These equations relate 15 dependent variables:
$p,\rho,\epsilon,\mu,\mu_b,\eta$ and the components of
$\textbf{u}, \textbf{J}$ and $\textbf{B}$. For the set of
equations, we use the following standard assumptions:

The kinematic viscosity coefficient, $\nu=\frac{\mu}{\rho}$, is
generally parameterized using the $\alpha$-prescription (Shakura-Sunyaev 1973),

\begin{equation}
\nu=\alpha c_sH,\label{nu1}
\end{equation}
where $H=\frac{c_s}{\Omega_k}$ is known as the vertical scale
height , $c_s=\sqrt{\frac{p}{\rho}}$ is the isothermal sound
speed and the dimensionless coefficient $\alpha$ is assumed to be
independent of r.  So, we introduce the parameter $\eta$ as the
magnetic diffusivity and insert it as a constant parameter in our
equations. Both the kinematic viscosity coefficient $\nu$ and the
magnetic diffusivity $\eta$ have the same units and are assumed to
be due to turbulence in the accretion flow. Thus it is physically
reasonable to express $\eta$ such as $\nu$ via the
$\alpha$-prescription of Shakura-Sunyaev (1973) as follows
(Bisnovatyi-Kogan \& Ruzmaikin 1976),
\begin{equation}
\eta=\eta_\circ c_sH.\label{eta1}
\end{equation}
To determine thermodynamical properties of the flow in the energy
equation, we require a constitutive relation as a
function of two state variables. Therefore we choose an equation for
the internal energy as $\epsilon=\frac{p}{\rho(\Gamma-1)}$ where
$\Gamma$ is the ratio of specific heats of the gas.

To satisfy $\nabla\cdot\textbf{B}=0$, we may introduce a convenient
functional form for the magnetic field. Owning to the axisymmetry,
the magnetic field can be written as
\begin{equation}
\textbf{B}=\textbf{B}_p(r,\theta)+B_\phi(r,\theta)\textbf{e}_\phi
\end{equation}
Angular momentum is expected to be carried away from the disk by magnetic stresses
along the externally given poloidal magnetic lines of force. In the case of dipole-type external field
it is transferred to the central accretor (Kaburaki. 2000).
The effect of magnetic diffusivity on magnetically driven mass
accretion was studied by Kaburaki (2000). They showed that
the effects of resistivity are that magnetic field lines do not
rotate with the same angular speed as the disk matter and thus it
suppresses the injection of magnetic helicity and
magneto-centrifugal acceleration. So, by neglecting the toroidal
component of the field, $B_\phi$, we can express the poloidal
component, $\textbf{B}_p$, in terms of a magnetic flux function
$\Psi(r,\theta)$:
\begin{equation}
\textbf{B}=\textbf{B}_p(r,\theta)=\frac{1}{2\pi}\nabla\times
\left(\frac{\Psi}{r\sin\theta}\textbf{e}_\phi\right)
\end{equation}

It is clear that the basic equations are nonlinear and we cannot
solve them analytically. Therefore, it is useful to have a simple
means to investigate the properties of solutions. We seek self-similar solution
for the above equations. In the next section we will present self-similar solutions
of these equations.

\section{Self-Similar Solutions}
To understand better the physical processes of our viscous-resistive ADAF accretion disks, we seek self-similar
solutions of the above equations. The self-similar method is
familiar from its wide applications to the full set of MHD
equations. The self-similar method is not able to describe the
global behavior of accretion flows, because no boundary condition
has been taken into account. However, as long as we are not
interested in the behavior of the flow near the boundaries, such
solutions are very useful.

Writing the equations in non-dimensional forms, that is, scaling
all the physical variables by their typical values, bring out
the non-dimensional variables. We can simply show that the
solutions of the following forms, satisfy the equations of our
model:
\begin{equation}
\rho(r,\theta)=\rho_\circ\rho(\theta)(r/r_\circ)^{-3/2},
\end{equation}
\begin{equation}
\ p(r,\theta)=p_\circ P(\theta)(r/r_\circ)^{-5/2},
\end{equation}
\begin{equation}
\ u_{\rm r}(r,\theta)= r\Omega_{\rm K}(r) U(\theta),
\end{equation}
\begin{equation}
\ u_{\rm \varphi}(r,\theta)=r\sin\theta\Omega_{\rm
K}(r)\Omega(\theta),
\end{equation}
\begin{equation}
\ B_{\rm
r}(r,\theta)=\frac{B_\circ}{2\pi\sin\theta}\frac{d\Psi(\theta)}{d\theta}(r/r_\circ)^{-5/4},
\end{equation}
\begin{equation}
\ B_{\rm
\theta}(r,\theta)=-\frac{3B_\circ\Psi(\theta)}{8\pi\sin\theta}(r/r_\circ)^{-5/4},
\end{equation}

where $\rho_\circ$, $p_\circ$, $B_\circ$ and $r_\circ$ provide convenient units with
which the equations can be written in non-dimensional forms. Substituting the above
solutions in the equations (10)-(16), we obtain a set of coupled
ordinary differential equations in terms of $\theta$.
\begin{displaymath}
\frac{dP}{d\theta}=\frac{3\alpha P}{2(1-\alpha
U)}\frac{dU}{d\theta}+\frac{3\rho KU}{16\pi^2 \beta_0 \eta_0 c_1 P
\Omega^{2} \sin^{2}\theta (1-\alpha U)}\frac{d\Omega}{d\theta}
\end{displaymath}
\begin{equation}
+\frac{\rho \Omega^2 \sin\theta \cos \theta}{c_1(1-\alpha U)}
\end{equation}

\begin{displaymath}
\frac{d^2 U}{d^2\theta}=-\frac{2.5}{\alpha}-U-\cot\theta
\frac{dU}{d\theta}-\frac{1}{P}\frac{dP}{d\theta}\frac{dU}{d\theta}
\end{displaymath}
\begin{equation}
+\frac{\rho}{c_1 \alpha P}(1-\frac{U^2}{2}+\Omega^2
\sin^2\theta)+\frac{2UK\rho}{\beta_0 \eta_0 \alpha c_1 \Omega
}(\frac{3}{8\pi \rho \sin\theta})^2\
\end{equation}

\begin{displaymath}
\frac{d\rho}{d\theta}=\frac{2}{5}\frac{c_{1}^{-1/2}P^{-1/2}\rho^{3/2}}{\phi_{s}}
[\frac{U(3\gamma-5)}{2(\gamma-1)}-\alpha
f(3U^2+(\frac{dU}{d\theta})^2+\frac{9}{4}\Omega^2
\end{displaymath}
\begin{displaymath}
\sin^2\theta+(\frac{d\Omega}{d\theta})^2
\sin^2\theta)]-\frac{2}{5}\frac{f
c_{2}\eta_{0}c_{1}^{-5/2}P^{-5/2}\rho^{5/2}
K}{16 \pi^3 \phi_{s}\Omega}(\frac{3U}{4 \eta_{0}\sin\theta})^2
\end{displaymath}
\begin{equation}
-2(1-\cot\theta)\rho +3\frac{\rho}{P}\frac{dP}{d\theta}
\end{equation}
\begin{equation}
\frac{d\Omega}{d\theta}=\frac{-A\pm \sqrt{{A^2+3B}}}{3}
\end{equation}

Finally, by definition $\beta_\circ=\frac{P_\circ}{B^2_\circ/8\pi}$, $\Omega
\Psi^2=K$, where $K$ is an arbitrary constant, so  $c_{\rm
1}=\frac{p_\circ}{\rho_\circ}\left(\frac{GM}{r_\circ}\right)^{-1}=\frac{2p_\circ}{\rho_\circ
u^2_{ff}}$ and $c_{\rm
2}=\frac{B^2_\circ}{\rho_\circ}\left(\frac{GM}{r_\circ}\right)^{-1}$.

In equation (30) we have the followings:
\begin{displaymath}
A=\frac{\Omega}{P}\frac{dP}{d\theta}+4\Omega\cot\theta,
\end{displaymath}
\begin{displaymath}
B=\left[\frac{9}{4}+(\frac{1}{\alpha}+\frac{3}{\eta_0})\frac{\rho
U}{c_1 P}\right]\Omega^2,
\end{displaymath}

Equations 27-30 constitute a system of ordinary non-linear
differential equations for the four self-similar variables
$\Omega, P, U  $\&$  \rho$.

There are many techniques for solving these nonlinear equations.
Analytical methods can yield solutions for some simplified
problems. But, in general this approach is too restrictive and we
have to use the numerical methods. Here, one can employ the
method of relaxation to the fluid equations (Press et al. 1992).
In this method we replace ordinary differential equations by
approximate finite-difference equations on a grid of points that
spans the domain of interest. The relaxation method determines
the solution by starting with a guess and improving it,
iteratively. Based on it, this system of equations can be solved
for all unknowns as a function of $\theta$, once we are given a
set of boundary conditions where constraints are placed on the
flow. The boundary conditions are distributed between the
equatorial plane, $\theta=\frac{\pi}{2}$ and the rotation axis,
$\theta=0$. We can use Narayan $\&$ Yi (1995)Boundary conditions
in both boundaries:
 the boundary conditions at $\theta=0$
\begin{equation}
\frac{dU}{d\theta}=\frac{d\Omega}{d\theta}=\frac{dP}{d\theta}=\frac{d\rho}{d\theta}=0~~~,~~~U=0,\rho=0,\label{condi1}
\end{equation}

and in this method the boundary conditions at
$\theta=\frac{\pi}{2}$ are:

\begin{equation}
\frac{dU}{d\theta}=\frac{d\Omega}{d\theta}=\frac{dP}{d\theta}=
\frac{d\rho}{d\theta}=0.\label{condi2}
\end{equation}
The boundary conditions on the above equations require that
variables are assumed to be regular at the endpoints. Also the
net mass accretion rate (9) provides one boundary condition for
$\rho$:
\begin{displaymath}
\int_{0}^{\frac{\pi}{2}}\rho(\theta)U(\theta)\sin\theta
d\theta=-\frac{1}{2}
\end{displaymath}

We obtain numerical solutions for the flows with fixed values of
$\eta_{0}=0.1$, $\Gamma=\frac{4}{3}$, $\alpha=0.01,0.05,0.1$
,$f=0.1,0.3,0.7$  and $\phi_{s}= 0.001,0.007,0.01$. We consider
$c_{1}=0.8$, $c_{2}=2\times10^{3}$ and  $\beta_{0}=0.01$
(Ghanbari , Salehi   $\&$ Abbassi 2007)

\section{Results}

\input{epsf}
\epsfxsize=3in \epsfysize=4.1in
\begin{figure}
\centerline{\epsffile{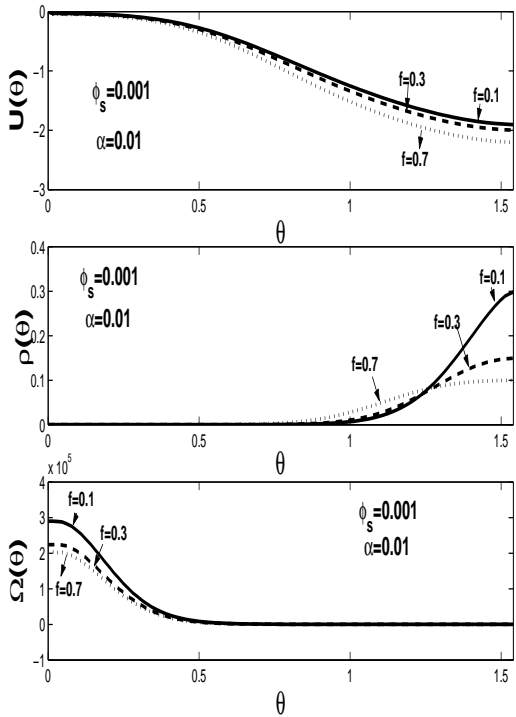}}
 \caption{The self-similar solutions of radial velocity $U(\theta)$ (top), density $\rho(\theta)$
 (middle) and angular velocity $\Omega(\theta)$ (bottom )as a function of polar angel $\theta$
  corresponding to $\Gamma=4/3$, $\eta_{0}=0.1$, $\beta_0=0.01$ and $f=0.1, 0.3, 0.7$ for $\alpha$=0.1
   and $\phi_{s}=0.001$}

 \label{figure1}
\end{figure}
\input{epsf}
\epsfxsize=3in \epsfysize=4.1in
\begin{figure}
\centerline{\epsffile{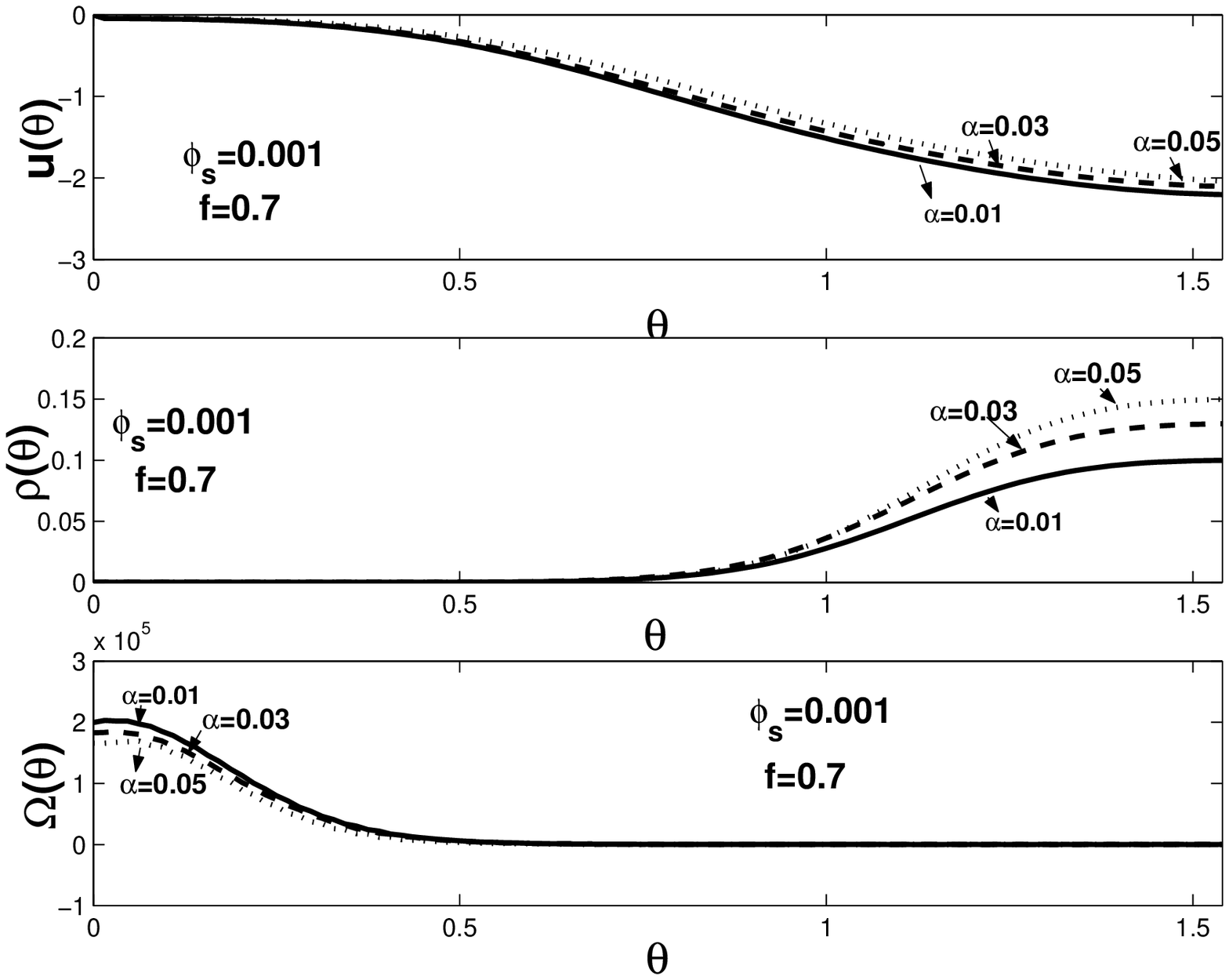}}
 \caption{ The self-similar solutions of radial velocity $U(\theta)$(top) and density $\rho(\theta)$
 (middle) and angular velocity $\Omega(\theta)$ (bottom) as a function of polar angel $\theta$
  corresponding to $\Gamma=4/3$, $\eta_{0}=0.1$, $\beta_0=0.01$ and $\alpha=0.01, 0.03, 0.05$ for
     $f=0.7$ $\phi_{s}=0.001$  }
 \label{figure2}
\end{figure}
\input{epsf}
\epsfxsize=3in \epsfysize=4.1in
\begin{figure}
\centerline{\epsffile{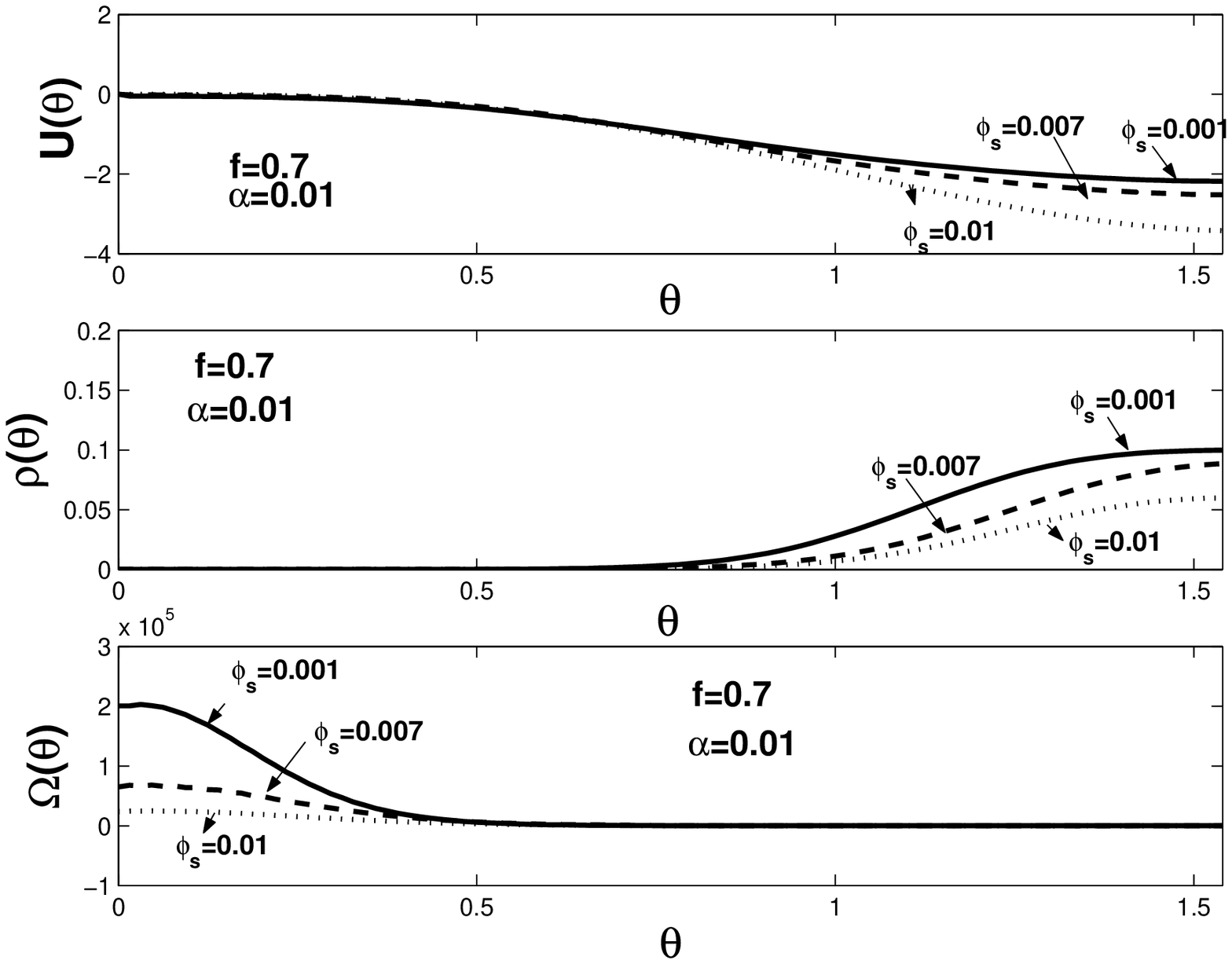}}
 \caption{The self-similar solutions of radial velocity $U(\theta)$(top) and density $\rho(\theta)$
 (middle) and angular velocity $\Omega(\theta)$ (bottom ) as a function of polar angel $\theta$
  corresponding to $\Gamma=4/3$, $\eta_{0}=0.1$, $\beta_0=0.01$ and $\phi_{s}=0.001, 0.007, 0.01$ for
     $f=0.7$ $\alpha=0.01$  }
 \label{figure3}
\end{figure}

\input{epsf}
\epsfxsize=3in \epsfysize=4.1in
\begin{figure}
\centerline{\epsffile{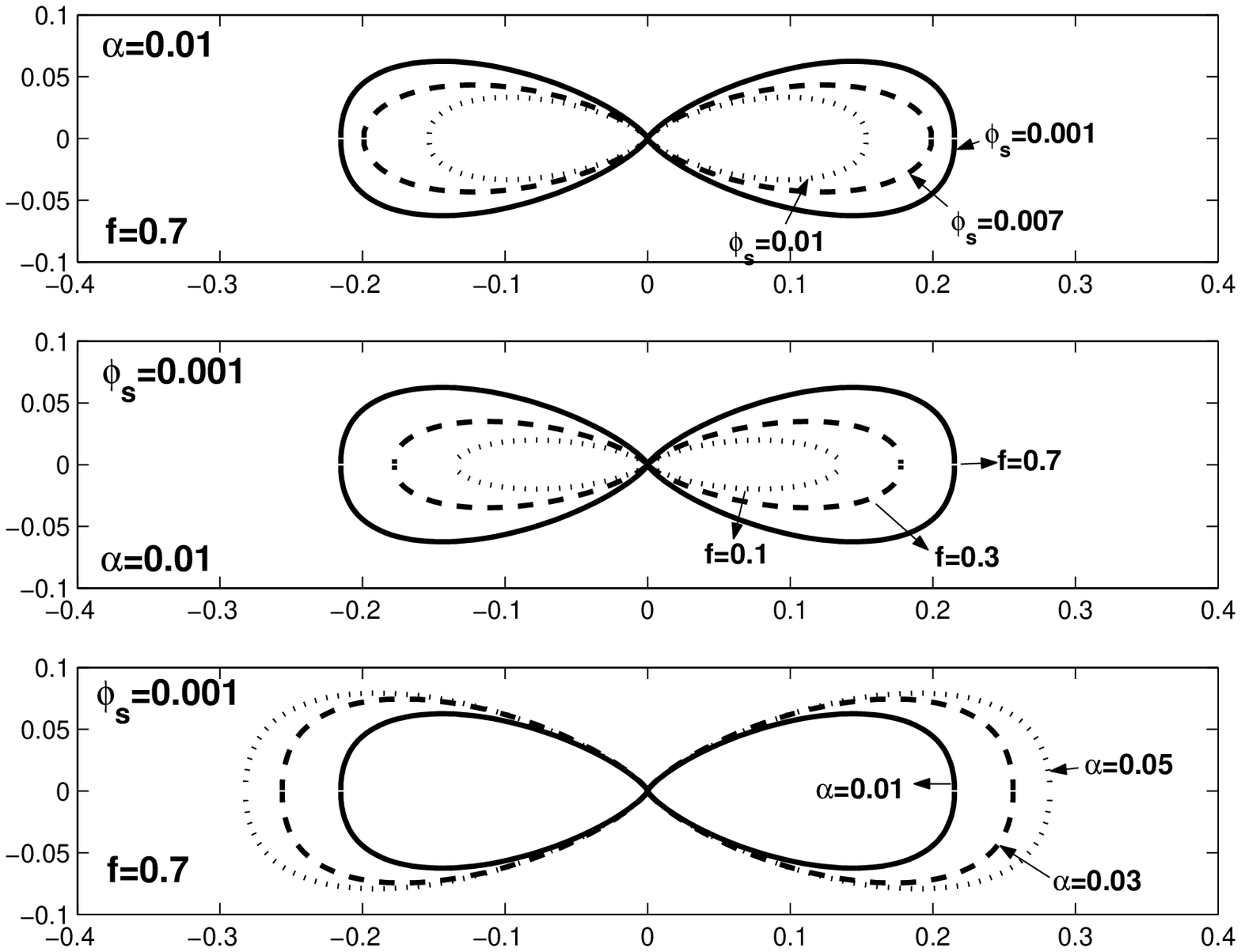}}
 \caption{isodensity contours for $\Gamma=4/3$, $\eta_{0}=0.1$, $\beta_0=0.01$ and $\phi_{s}=0.001, 0.007, 0.01$ for
     $f=0.7$ $\alpha=0.01$(top) and $f=0.1, 0.3, 0.7$ for $\alpha$=0.01
   and $\phi_{s}=0.001$(middle) and $\alpha=0.01, 0.05, 0.1$ for
     $f=0.7$ $\phi_{s}=0.001$ (bottom) }
 \label{figure4}
\end{figure}

We have obtained numerical solutions of equations (27)-(30) for
a variety of viscosity, $\alpha$ , advection, $f$ and the thermal conduction $\phi_{s}$ parameters. The
three panels in Figure 1 show the variations of various dynamical quantities in terms of polar angle $\theta$ for with a
fixed values of viscosity and thermal conduction parameters
with a sequence of increasing advection parameter $f$. The top
panel displays the dimensionless radial velocity $U(\theta)$.
 $U(\theta)$ is zero at $\theta=0$ (this is a boundary condition)
and maximum at $\theta=\frac{\pi}{2}$ . \textbf{Thus, the inflow velocity
reaches its maximum in the equatorial plane and vanishes along the polar axis.}
As expected , the velocity is sub-Keplerian.
The middle panel shows the density profile $\rho(\theta)$ of the
solutions. The density contrast in the equatorial and polar
regions increases with a decrease in the advection parameter $f$.
For a given $\alpha$ and $\phi_s$ , solutions with small values of
$f$ behave like standard thin disks, as might be expected, these solutions correspond to $f\rightarrow 0$ and so
a small fraction of energy would be advected. In the opposite, $f\rightarrow 1$, advection-dominated limit, our solutions describe nearly
spherical flows which rotate far below the Keplerian velocity.
The bottom panel shows the profile of the angular velocity
$\Omega(\theta)$. $\Omega(\theta)$ decreases increase in
the advection in the accretion disks. We find that in
the inner boundary, $U(\theta)$ is essentially independent of advection
parameter $f$. But in intermediate values of $\theta$, the radial velocity is
modified by $f$; in the Shadmehri's solutions (2004), two distinct regions in the
$U(\theta)$ profile could be recognized. The bulk of accretion
occurs from equatorial plane at $\theta=\frac{\pi}{2}$ to $\theta=\theta_s$, in which the radial velocity
is zero. While in Narayan $\&$ Yi solutions, there is no zero inflow
in $0<\theta<\frac{\pi}{2}$. Our solutions show that in any given
$\theta$ the radial velocity is nonzero and when we increase the
advection parameter the radial velocity will be increased.

Here we may comment about the fact that when the advection parameter,
$f$, goes to zero, our disk dose not correspond to a globally cooling flow,
because of the appearance of thermal conduction term in the energy transport equation.
When $f$ goes to 1 our disks are not fully advective, because some part of the energy
generated by viscosity will be transported by thermal conduction.

Figure 2 displays the behavior of radial and angular velocities
and density profile for different values of the viscosity
parameter for fixed advection and thermal conduction parameters.
We find that the value of viscous parameter, $\alpha$, affects
quantitatively (but not qualitatively) on the dynamical variables
of the accretion flow. For the larger value of the viscous
parameter, the radial inflow decreases and the
density would be increased overally, which is compatible with the results presented
by Ghanbari et al (2007).

Three panels in Figure 3 show, for fixed advection and viscosity
parameters with a sequence of thermal conduction parameters,:
(1) The radial velocity increases with an increase in
the thermal conduction parameter (2) The density profile
increases with a decrease in the thermal conduction parameter
(3) The angular velocity decreases with an increase in the
$\phi_{s}$ in the accretion discs.

In Fig. 4 we display the isodensity contours in the meridional plane. The top, middle and bottom panels
display the isodensity profiles for different values of conduction, advection and viscous parameters, respectively.
 The panels in figure 4 show that the disk seems to be
thick. Solution with the same $f$ but for different values of $\alpha$
are distinguishable from one another. By adding viscous parameter the geometrical shape of the disc becomes thick
more and more. This advection-dominated solutions have very similar properties to the approximated
solution derived by Narayan \& Yi (1994), Ghanbari et ql (2007) and Shadmehri (2004).

The overall structure of the dynamical variables remain very
close to the original 2D ADAF solutions of Narayan \& Yi (1995).
This solution is denser close to the equator than at the pole,
and it is some thing like a "thin disk" surrounded by a hot
coronal atmosphere.

As the values of $\phi_s$ are increased, the solutions
substantially start to deviate from standard solutions, with
faster radial flow and rotating more slowly (becomes more
pressure supported) in the equator. By increasing $\phi_s$, density reduces in the
equatorial plane, and the density profile
becomes more uniform . It gradually approaches to the spherically
symmetry. This behavior of the solutions are shared by original
ADAFs solutions and Ghanbari et al.'s work  (2007)

\section{SUMMERY AND CONCLUSION }

The main aim of this investigation was to obtain an axi-symmetric
self-similar advection-dominated solution for viscous-resistive
accretion flow with the poloidal magnetic field in the presence of
thermal conduction. Using the basic equations of fluid dynamics in
spherical polar coordinates $(r,\theta,\varphi)$, we have found
the self-similar solutions for thick discs to derive a set of
coupled differential equations that govern the dynamics of the
system. We have then solved the equations using the relaxation
method by considering boundary conditions and using the
$\alpha$-prescription (Shakura \& Sunyaev, 1973) in order to
extract some of the similarity functions in terms of the polar
angle $\theta$.

We showed that the radial and rotational velocities are well
below the Keplerian velocity. The Bulk of accretion with nearly
constant velocity occurs in the regions which extend from
equatorial plane to a given $\theta$ which highly depends on
advection parameter $f$. In a non-advective regime, low $f$, we
have a standard thin accretion disk, but for a high $f$ the accretion
is nearly spherical.

It is difficult to evaluate the precise picture of radiative
inefficient accretion flow in the presence of thermal conduction
with a self-similar method. But this method can reproduce overall
dynamical structure of the disks with a set of given physical
parameters. Even though, conduction heats up  the accretion
flows locally, the reduced density resulting from the larger
inflow velocity, leads a net decrease in the expected level of
free-free emissions. The very steep dependence of synchrotron
emissions on the electron temperature (e.g. Mahadevan \& Quataret 1997),
suggests that hotter solutions (with conduction) maybe
 more efficient radiatively ( Tanaka \& Menou 2006). From
self-similar solutions alone, we can determine how the global
structure of the flow can be affected by thermal conduction.
For small enough values of $\phi_s$ the solutions remain very close
to the Ghanbari et al. (2007) solutions. The main difference between their solution with
standard solution of 2D ADAFs of Narayan \& Yi (1995a) is the presence of dipolar
magnetic field and its correspond resistivity. In our case, we add extra physics by adding the
thermal conduction as a mechanism for energy transport. By adding the thermal conduction
the solution start deviating substantially from the original ADAFs, with faster radial inflow at the
equator. These results well agree with Tanaka \& Menou (2006).

The presence of magnetic field with the poloidal configuration will affect the role of
thermal conduction. Compare to non-magnetic field solution, (Tanaka \& Menou 2006)
in our case the existence of magnetic resistivity can produce more energy to be advected.
The B-field configuration can also affect the energy transportation along the accretion disks.
However, the main aim of this work is to study the quasi-spherical magnetized flow, directly by
solving the relevant MHD equations. Although, we have made some simplifications
in order to treat the problem analytically, our self-similar solutions show that
the input parameters, such as thermal conduction and viscose parameters, magnetic field and its resistivity can really change typical behavior of the physical quantities of the ADAF disks. Of course, our self-similar solutions are too simple to make any comparison with observations. But, we think
one may relax self-similarity assumptions and solve the equations of the model numerically. This kind of similarity solutions can greatly facilitate testing and interpreting the results.

\section{ACKNOWLEDGMENTS}
We are grateful to the referee for a very careful reading of the manuscript and for his/her suggestions, which have helped us improve the presentation of our results.

\end{document}